\documentclass[conference]{IEEEtran}
\IEEEoverridecommandlockouts
\usepackage{cite}
\usepackage{amsmath,amssymb,amsfonts}
\usepackage{algorithmic}
\usepackage{graphicx}
\usepackage{textcomp}
\usepackage{xcolor}
\usepackage{multirow}
\usepackage{booktabs}
\usepackage{tikz}
\newcommand{\tikzxmark}{%
\tikz[scale=0.23] {
    \draw[line width=0.7,line cap=round] (0,0) to [bend left=6] (1,1);
    \draw[line width=0.7,line cap=round] (0.2,0.95) to [bend right=3] (0.8,0.05);
}}
\newcommand{\tikzcmark}{%
\tikz[scale=0.23] {
    \draw[line width=0.7,line cap=round] (0.25,0) to [bend left=10] (1,1);
    \draw[line width=0.8,line cap=round] (0,0.35) to [bend right=1] (0.23,0);
}}

\def\BibTeX{{\rm B\kern-.05em{\sc i\kern-.025em b}\kern-.08em
    T\kern-.1667em\lower.7ex\hbox{E}\kern-.125emX}}
\begin{document}

\title{TDFNet: An Efficient Audio-Visual Speech Separation Model with Top-down Fusion}

\author{\IEEEauthorblockN{Samuel Pegg$^{1, \dagger}$, Kai Li$^{1, \dagger}$, Xiaolin Hu$^{1,2,3,*}$}
\IEEEauthorblockA{1. \textit{Department of Computer Science and Technology, Institute for AI} \\
\textit{BNRist, Tsinghua University, Beijing 100084, China}\\
2. \textit{Tsinghua Laboratory of Brain and Intelligence (THBI)}, \\ \textit{IDG/McGovern Institute for Brain Research, Tsinghua University, Beijing 100084, China} \\
3. \textit{Chinese Institute for Brain Research (CIBR), Beijing 100010, China} \\
axh.2020@tsinghua.org.cn, lk21@mails.tsinghua.edu.cn, xlhu@tsinghua.edu.cn}
\thanks{$\dagger$ Samuel Pegg and Kai Li have contributed equally to this work.}
\thanks{* Corresponding author.}
}

\maketitle

\begin{abstract}
    Audio-visual speech separation has gained significant traction in recent years due to its potential applications in various fields such as speech recognition, diarization, scene analysis and assistive technologies. Designing a lightweight audio-visual speech separation network is important for low-latency applications, but existing methods often require higher computational costs and more parameters to achieve better separation performance.
    In this paper, we present an audio-visual speech separation model called Top-Down-Fusion Net (TDFNet), a state-of-the-art (SOTA) model for audio-visual speech separation, which builds upon the architecture of TDANet, an audio-only speech separation method. TDANet serves as the architectural foundation for the auditory and visual networks within TDFNet, offering an efficient model with fewer parameters. On the LRS2-2Mix dataset, TDFNet achieves a performance increase of up to 10\% across all performance metrics compared with the previous SOTA method CTCNet. Remarkably, these results are achieved using fewer parameters and only 28\% of the multiply-accumulate operations (MACs) of CTCNet. In essence, our method presents a highly effective and efficient solution to the challenges of speech separation within the audio-visual domain, making significant strides in harnessing visual information optimally. 
\end{abstract}


\section{Introduction}

\begin{figure}[ht]
  \centering
  \includegraphics[width=0.8\linewidth]{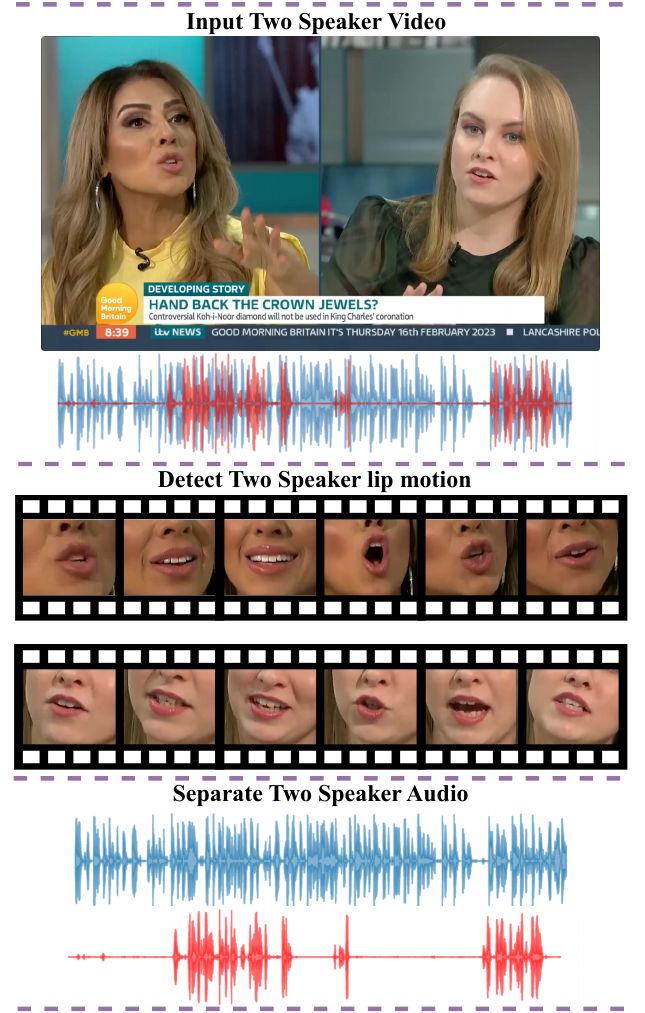}
  \caption{
  Audio-visual speech separation process. From top to bottom: video frames and mixed inputs, cutting out lip regions from the video, TDFNet speech separation results.
  }
  \label{fig:1}
\end{figure}

\begin{figure*}[ht]
  \centering  
  \includegraphics[width=0.95\linewidth]{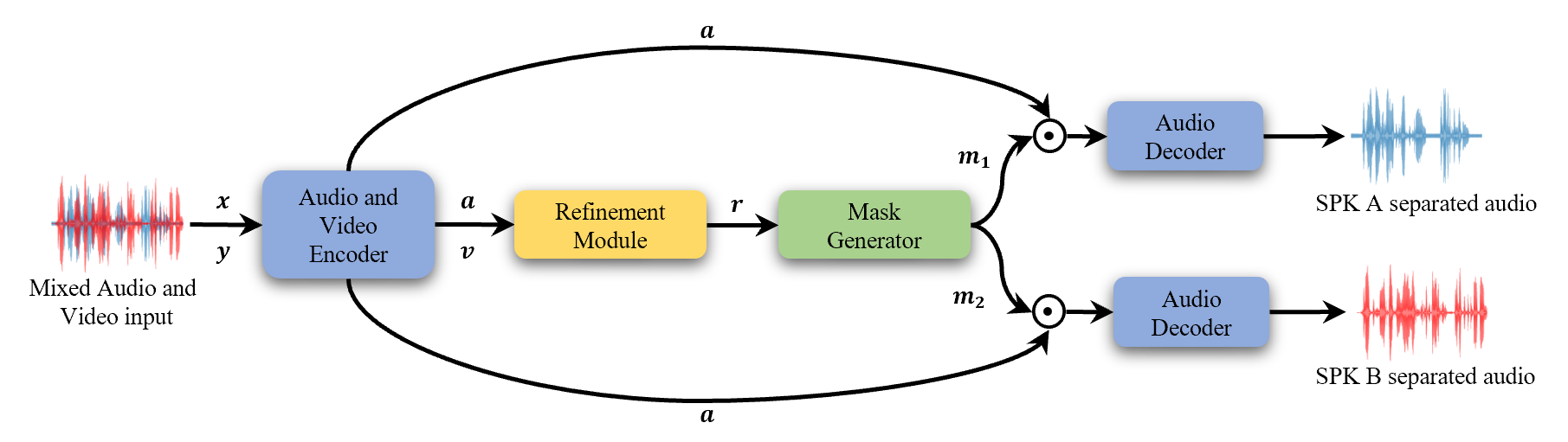}
  \caption{TDFNet separation pipeline. The audio and video inputs $\pmb{x}$ and $\pmb{y}$ are encoded by $E_a$ and $E_v$ respectively to produce the feature maps $\pmb{a}$ and $\pmb{v}$, which are sent to the refinement module $R$ to be fused and then further processed. The mask generator $M$ then takes these refined features $\pmb{r}$ and generates masks $\pmb{m}_i$, which are multiplied by the encoded audio input $\pmb{a}$ in tern to produce a separation. Finally, the decoder decodes each of the separated encoded audios. The figure above uses $n_{spk}=2$ speakers.}  
  \label{fig:TDFNet}
\end{figure*}

Speech separation is the process of extracting distinct audio streams from an audio recording containing one or more speakers \cite{unified_framework_for_speech_separation}. Consider a scenario where we have a microphone positioned in the center of a room capturing the voices of two individuals, A and B. These speakers may talk simultaneously or one after the other, with varying volume levels and distances from the microphone. It is crucial to account for all these factors. The objective of a speech separation model is to split the audio recording into two separate streams, each containing the audio from a single speaker. Ideally, one output stream would exclusively contain the voice of speaker A or B. 

Speech separation is commonly referred to as the ``cocktail party problem" \cite{cocktail_party_problem, cocktail_party_phenomenon}. At social gatherings, such as cocktail parties, our natural inclination is to concentrate on a particular individual while filtering out the surrounding conversations. Over the past decade, High quality speech separation has become increasingly crucial due to the widespread adoption of automated systems and voice assistants such as Apple's Siri and home devices such as Amazon Alexa or Google Home \cite{KimMCHNSB17, acoustic_modeling_google_home, 8462269,sharif2020smart, haeb2019speech, martinek2020voice}. 

Methods based on architectures designed for modeling sequences, such as Recurrent Neural Networks (RNNs) \cite{dprnn}, or extracting local patterns such as Convolutional Neural Networks (CNNs) \cite{Conv-TasNet}, have proven to be adept at handling the complexities in speech signals. However, methods relying on auditory signals alone, known as audio-only speech separation (AOSS) methods, have become quite saturated in recent years, with new models showing only incremental performance increases. One notable avenue to bolster the robustness of speech separation is to integrate multi-modal information \cite{smith2005development, cappe2009multisensory}. For humans, the integration of auditory information (speech signals) and visual stimuli (such as lip movements) fundamentally alters our perception of language \cite{pallas1990visual, morrill2018visual, donishi2011sub}. 

Audio-visual speech separation (AVSS) is similar to speaker extraction \cite{xu2020spex,ge2020spex+,wang2018voicefilter}, which uses one speaker's voiceprint to create a separation. However, whereas voice extraction requires enrolling a target speaker in advance and gathering their voiceprint, visual information can be entirely captured in real-time, as shown in Figure \ref{fig:1}. Visualvoice \cite{VisualVoice} and CTCNet \cite{CTCNet} both use an encoder-decoder structure to improve the separation performance, but they do not fully exploit this structure. In addition, integrating audio and visual information can be computationally expensive, making them less practical for real-time applications.

In order to efficiently extract and integrate features from different modalities, we propose a novel AVSS model called \textbf{TDFNet}. It uses encoded audio and visual information as inputs for a series of TDANet-based blocks\cite{TDANet}, called TDFNet Blocks. These blocks build a hierarchical structure where the lower levels have higher temporal resolution. Information from different temporal resolutions moves freely, and is aggregated and fused with the visual information. We evaluated TDFNet extensively on the LRS2 benchmark dataset to show that it outperforms various competing baseline models and achieves state-of-the-art performance using fewer parameters and with a lower computational cost.

\section{Related Work}

\subsection{Audio-Only Speech Separation}

\label{sec:aa_speech_sep}

Initially, speech separation employed the time-frequency (TF) representation of the mixed audio, estimated from the waveform using the short-time Fourier transform (STFT). Pioneering methods utilized matrix factorization\cite{matrix_factorization} and heuristic techniques\cite{speech_separation_and_recognition_challenge} to cluster the TF bins of each speaker. However, the performance of these models was either poor or speaker-dependent. With the development of deep learning and the introduction of permutation invariant training (PIT\cite{yu2017permutation}) to solve the permutation problem, the speech separation space has become increasingly competitive. Researchers soon migrated from the TF representation of audio to a time domain only representation using a convolutional encoder, resulting in high performance methods such as DualPathRNN \cite{dprnn}, Sepformer \cite{subakan2021attention_sepformer}, Wavesplit \cite{zeghidour2021wavesplit}, etc. Recently, the multi-scale speech separation model AFRCNN \cite{AFRCNN} was proposed as a balance between efficiency and separation performance. TDANet \cite{TDANet} adds top-down attention \cite{mizokuchi2023alpha, song2023total} to AFRCNN's encoder-decoder architecture to obtain state-of-the-art performance with a greatly reduced computational effort. Both of these models were inspired by the brain.

Some researchers suggest that the datasets used in the audio-only setting are often idealized, lacking reverberation, interfering background sounds or random noise. A recent study \cite{TDANet} found that many top models become average when faced with these more challenging datasets. Furthermore, audio-only models encounter significant challenges when confronted with scenarios involving three or more speakers\cite{deep_clustering_discriminative_embeddings_for_segmentation_and_separation}, or when the number of speakers is unknown\cite{xu2020spex,ge2020spex+,wang2018voicefilter}. 

\subsection{Audio-Visual Speech Separation}

The audio-visual speech separation field combines audio-only speech separation with video data to improve performance on noisy and challenging datasets. Recent neuroscience studies have demonstrated that the human brain effectively addresses the cocktail party problem by leveraging additional visual cues from the eyes\cite{bhx235, fnins_2019_00451}. Building upon this notion, some researchers\cite{CaffNet_C, Time_Domain_Audio_Visual_Speech_Separation, VisualVoice, the_conversation_Deep_Audio_Visual_Speech_Enhancement, A_cappella_Audio_visual_Singing_Voice_Separation, Looking_to_listen_at_the_cocktail_party,pan2021muse, lin2023av_sepformer} have endeavored to incorporate visual information into the paradigm, aiming to enhance the quality of audio separation. These efforts have culminated in the current state-of-the-art for audio-visual speech separation, namely CTCNet\cite{CTCNet}. However, CTCNet, due to the complexity of its multi-scale fusion operations, introduces a substantial amount of computation which limits its applicability in real-world scenarios. To address this, we utilize TDANet's architecture for our visual and auditory feature extraction networks, which employ progressive multi-scale fusion, effectively reducing the computational cost of the multi-scale fusion.

\section{TDFNet}

\begin{figure*}[ht]
    \centering
    \includegraphics[width=0.75\linewidth]{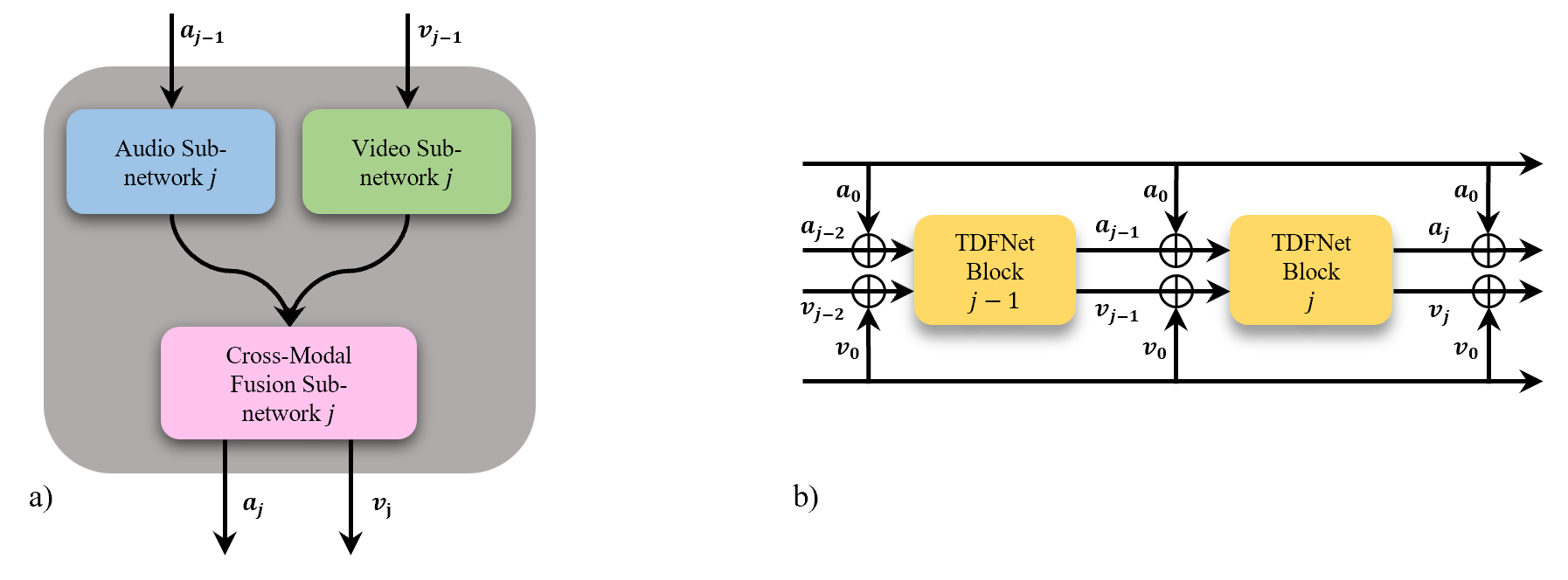}
    \caption{The $j$\textsuperscript{th} TDFNet Block. (a) The internals of a single TDFNet block. (b) The stacked TDFNet blocks with residual connection from the first iteration.}
    \label{fig:refinement_module_structure}
\end{figure*}

For a black and white 25 fps video containing $n_{spk}$ speakers, the inputs for our AVSS pipeline are: a sequence of video frames $\pmb{y}\in \mathbb{R}^{1 \times L_v \times H_{in} \times W_{in}}$ containing the lip movements of the desired speaker, and the mixed audio stream $\pmb{x}\in \mathbb{R}^{1\times L_a}$, where $L_v$, $H_{in}$, $W_{in}$ and $L_a$ denote the number of video frames, image height, width, and the length of audio, respectively. We assume that the mono-aural (single channel) audio of the video $\pmb{x}$ consists of linearly superimposed voices $\pmb{s}_i\in \mathbb{R}^{1\times L_a}$ for $i \in [1,n_{spk}]$ such that 
\begin{equation}
    \pmb{x} = \pmb{\epsilon} + \sum_{i=1}^{n_{spk}} \pmb{s}_i,
\end{equation}
where $\pmb{\epsilon} \in \mathbb{R}^{1\times L_a}$ is some presence of background noise. 

We propose TDFNet, which consists of five main modules: a video encoder (Section \ref{sec:1d_video_encoder}), an audio encoder (Section \ref{sec:1d_audio_encoder}), a refinement module (Section \ref{sec:refinement_module}), a mask generator (Section \ref{sec:mask_generator}) and an audio decoder (Section \ref{sec:decoder}). The pipeline of TDFNet is briefly described below (see Figure \ref{fig:TDFNet}).

\begin{enumerate}
    \item 
        The raw audio and visual signals are sent to respective encoders to obtain rich feature maps. We denote the audio encoder $E_a(\cdot)$ and the video encoder $E_v(\cdot)$:
        \begin{align*}
            \pmb{a} &= E_a(\pmb{x}), \quad  \pmb{a} \in \mathbb{R}^{C_a\times T_a }, \\
            \pmb{v} &= E_v(\pmb{y}), \quad  \pmb{v} \in \mathbb{R}^{C_v \times T_v },
        \end{align*}
        where the $C_a$ and $C_v$ denote the dimensions of visual and auditory features. Historically, large values for $C_a$ and $C_v$ are used as this leads to a more detailed feature representation. However, this increases the computational cost of the refinement module. In order to mitigate this cost while maintaining performance, we design a ``bottleneck" by using fewer channels $B_a \leq C_a$ and $B_v \leq C_v$. This is achieved by using 1D convolutional layers $F_a(\cdot)$ and $F_v(\cdot)$ with kernel size $1$. This can be written
        \begin{equation*}
            \pmb{a}' = F_a(\pmb{a}), \quad \pmb{v}' = F_v(\pmb{v}).
        \end{equation*}
    \item
        The refinement module $R(\cdot,\cdot)$, takes the audio and visual information and uses the combined information to generate a refined feature map:
        \begin{equation*}
            \pmb{r} = R(\pmb{a}',\pmb{v}'), \quad \pmb{r} \in \mathbb{R}^{B_a \times T_a }
        \end{equation*}
        We can see that the output dimensions match the dimensions of the \textit{audio} input. This is because the refinement module first fuses the visual information \textit{into} the audio information, and then further processes these combined multimedia features. This will be explained in Section \ref{sec:refinement_module}.
    \item
        This refined feature map is used to generate masks for each speaker. Let $M(\cdot)$ denote a mask generating function. 
        \begin{equation*}
            \{\pmb{m}_1, \dots , \pmb{m}_{n_{spk}}\} = M(\pmb{r}),
        \end{equation*}
        where $\pmb{m}_i \in [0,1]^{C_a\times T_a }$ for all $i \in [1,n_{spk}]$ and $[0,1]$ is the range of real numbers from 0 to 1. Note that each mask has the same dimensions as the encoded audio $\pmb{a}$.
    \item
        The encoded audio input is multiplied element-wise by each mask in turn, resulting in the separated encoded audio for each speaker,
        \begin{equation*}
            \pmb{z}_i = \pmb{a} \odot \pmb{m}_i, \quad \pmb{z}_i \in \mathbb{R}^{C_a\times T_a } \; \forall i\in [1,n_{spk}],
        \end{equation*}
        where $\odot$ represents element-wise multiplication.
    \item
        The separated audio feature map for each speaker is given to the decoder $D_a(\cdot)$. The decoder returns the separated audios for each speaker as a waveform,
        \begin{equation*}
            \hat{\pmb{s}}_i = D_a(\pmb{z}_i),  \quad \hat{\pmb{s}}_i \in  \mathbb{R}^{1 \times L_a} \; \forall i\in [1,n_{spk}].
        \end{equation*}
\end{enumerate}

\subsection{Video Encoder}
\label{sec:1d_video_encoder}

The video encoder tales the gray-scale video frames $\pmb{y}$ and uses the pre-trained lip-reading model used in CTCNet\cite{CTCNet}, called CTCNet-Lip, to extract visual features. It consists of a backbone network for extracting features from the image frames and a classification sub-network for word prediction. The backbone network includes a 3D convolutional layer and a standard ResNet-18 network. The frames are convolved with $P$ 3D-kernels with size $5 \times 5 \times 1$ and spatial stride size $2 \times 2 \times 1$ to obtain a rich feature map. Each feature map is passed through the ResNet-18 network, and the resulting feature maps are passed to the classification sub-network as words for word prediction. After lip-reading pre-training, the backbone network of the CTCNet-Lip is fixed for extracting visual features, and the classification sub-network is discarded. We encourage readers to read the original paper for more details.

\subsection{Audio Encoder}
\label{sec:1d_audio_encoder}

The audio encoder $E_a$ takes the input audio $\pmb{x}$ and produces a feature map using a 1D-convolution with $C_a$ output channels and kernel size $K_a$, followed by global layer normalization (gLN) \cite{Conv-TasNet} and ReLU activation. Since a stride greater than one is used, this technically changes the $T_a$ dimension to a smaller value, compressing the audio, but for simplicity we do not change the notation.  

\subsection{Refinement Module}
\label{sec:refinement_module}

The refinement module aims to take the audio and the video embeddings, fuse them together and then further process and refine the resulting multi-modal features. This way, we combine the audio and visual information in order to increase the separation performance. As is seen several times in the audio-only domain\cite{TDANet, AFRCNN}, we take an iterative approach. In total, we use $R_a$ iterations with fusion occurring $R_f$ times, where $R_f\leq R_a$.

The refinement module consists of three networks that work together in order to produce the best possible amalgamation of data:
\begin{itemize}
    \item $\alpha_j$, the \textbf{audio sub-network} at iteration $j$.
    \item $\beta_j$, the \textbf{video sub-network} at iteration $j$.
    \item $\gamma_j$, the \textbf{cross-modal fusion sub-network} at iteration $j$.
\end{itemize}
These three networks do not change the dimensions of their inputs, which is key for their inter-compatibility. For the \textit{fusion} iterations, for $j \in [1, R_f]$, one iteration consists of all three modules working together for form a TDFNet block. For the subsequent audio-only iterations, for $j \in [R_f,R_a]$, the TDFNet block will will consist of \textit{only} the audio sub-network.

Firstly, let $\pmb{a}_{j-1}$ and $\pmb{v}_{j-1}$ be the inputs for $\alpha_j$ and $\beta_j$, where the ``$0^{\text{th}}$" iteration is simply taking the outputs of the audio and video ``bottleneck" convolutions:
\begin{equation}
    \pmb{a}_{0} := \pmb{a}' \in\mathbb{R}^{B_a\times T_a}, \quad \pmb{v}_{0} := \pmb{v}' \in\mathbb{R}^{B_v\times T_v}.
\end{equation}

The first iteration is the most simple. The audio features are passed to the audio sub-network and the video features are passed to the video sub-network. The outputs are then passed to the fusion module which fuses the audio information with the visual information, and the visual information with the audio information. In the proceeding layers, residual connections are added to improve training.
\begin{align}
    (\pmb{a}_{1},\; \pmb{v}_{1}) &= \gamma_1\left(\alpha_1(\pmb{a}_{0}),\; \beta_1(\pmb{v}_{0})\right) \\ 
    \label{eq:j2_to_jrf}
    (\pmb{a}_{j}, \; \pmb{v}_{j})  &= \gamma_j \left(\alpha_j ( \pmb{a}_{j-1} + \pmb{a}_{0}), \; \beta_j(\pmb{v}_{j-1} + \pmb{v}_{0}) \right)    
\end{align}  
for $j\in [2,R_f]$. The interactions can be seen in Figure \ref{fig:refinement_module_structure}.

After $R_f$ fusion iterations, we deem the video and audio signals to be sufficiently `fused' together into the audio signal $\pmb{a}_{R_f} \in \mathbb{R}^{B_a \times T_a }$. For subsequent iterations, the signal is continuously refined using \textit{only} the audio sub-network. This simplifies Equation \ref{eq:j2_to_jrf} to the one below.
\begin{equation}
    \pmb{a}_{j} = \alpha_j\left(\pmb{a}_{j-1} + \pmb{a}_{0}\right),
\end{equation}
for $j\in [R_f+1,R_a]$.

\subsubsection{Audio and Video Sub-Networks}
\label{sec:avsubnets_briefMention}

Both $\alpha_j$ and $\beta_j$ are defined by the same architecture inspired by TDANet\cite{TDANet} and consists of three main sections:
\begin{enumerate}
    \item Bottom-up Down-sampling
    \item Recurrent Operator
    \item Top-down Fusion
\end{enumerate}
The details of these sections will be covered in Section \ref{sec:avsubnet_chapter}. We define
\begin{align*}    
    \pmb{X}_j &= \alpha_j \left( \pmb{a}_{j-1} + \pmb{a}_{0}\right) &\in \mathbb{R}^{B_a\times T_a},
    \\
    \pmb{Y}_j &= \beta_j\left(\pmb{v}_{j-1} + \pmb{v}_{0}\right) &\in \mathbb{R}^{B_v\times T_v},
\end{align*}
as the outputs of the audio and video sub-networks, and hence the two inputs for the cross-modal fusion sub-network.

\subsubsection{Cross-Modal-Fusion Sub-network}

This module is in charge of fusing the audio features into the video features, and the video features into the audio features. Let $\kappa$ define a 1D-convolution with kernel size $1$ followed by a gLN layer, and let $\phi$ denote nearest neighbor interpolation. $||$ is the concatenation operation acting along the channel dimension. Wielding these definitions, we can define the cross-modal fusion sub-network $\gamma_j$ that returns two outputs, and hence we can write the full expression for the $j$\textsuperscript{th} iteration as
\begin{equation}
    \gamma_j(\pmb{X}_j,\pmb{Y}_j) = \left(\pmb{a}_j, \; \pmb{v}_j \right),
\end{equation}
where
\begin{align*}
    \pmb{a}_j &= \kappa \left( \pmb{X}_j || \phi \left(\pmb{Y}_j  \right) \right) &\in \mathbb{R}^{B_a\times T_a}, \\
    \pmb{v}_j &= \kappa \left( \pmb{Y}_j || \phi \left(\pmb{X}_j  \right) \right) &\in \mathbb{R}^{B_v\times T_v}.
\end{align*}

As we can see, the video features are interpolated to match the audio dimensions: 
\begin{equation*}
    B_v \times T_v  \Longrightarrow B_v \times T_a  .
\end{equation*}
The output is concatenated with the audio features and then passed through a convolution layer to take the dimensions back to the dimensions of the audio input: 
\begin{equation*}
    (B_a + B_v) \times T_a  \Longrightarrow B_a \times T_a .
\end{equation*}
The video fusion result is similar.

\begin{figure*}[ht]
    \centering
    \includegraphics[width=.95\linewidth]{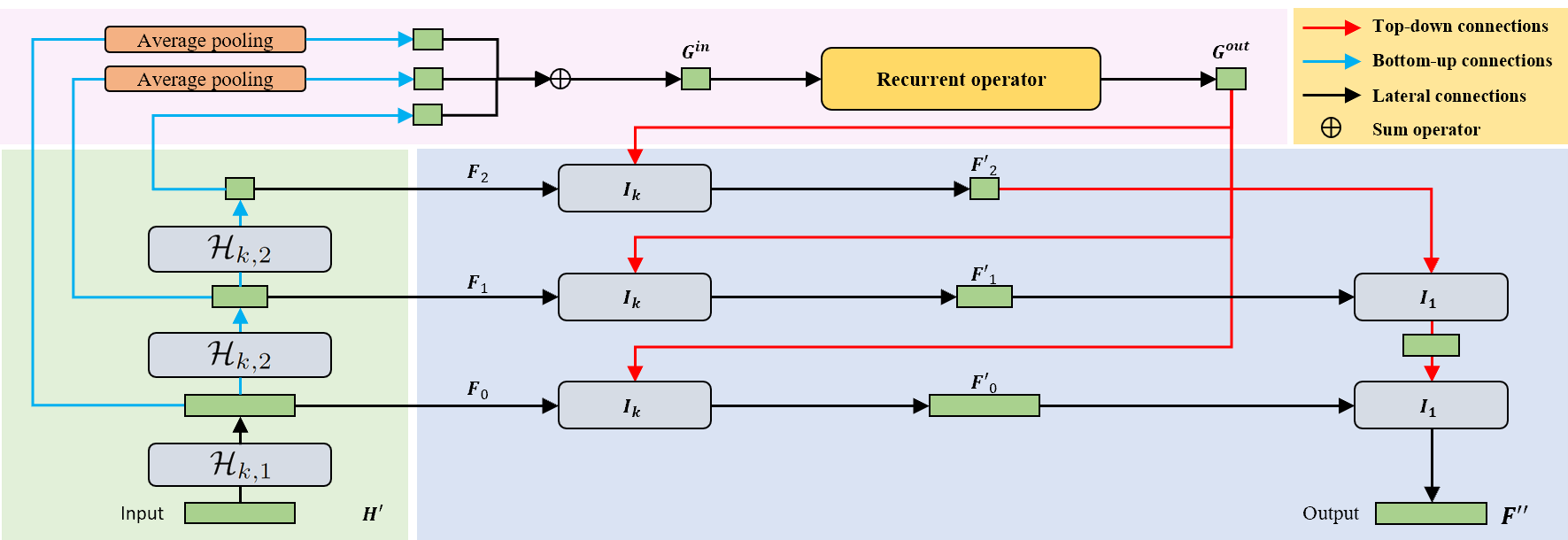}
    \caption{The core architecture of the audio and video sub-networks. The input is either the audio or the visual features, reduced to the hidden dimension $D$. The bottom-up down-sampling in the green block uses consecutive convolutions with stride 2 to compress the data to increasingly small temporal resolutions. The recurrent operator in the pink block fuses the information to formulate a global perspective. The top-down fusion in the blue block combines the global information at different temporal resolutions, and then fuses them all together into a single feature map.}
    \label{fig:TDF_internals}
\end{figure*}

\subsection{Mask Generator}
\label{sec:mask_generator}

The mask generator is tasked with taking the output of the refinement module $\pmb{r}$ and transforming it into a set of masks $\pmb{m}_i$ for $i\in [1,n_{spk}]$ and multiplying each mask with the encoded audio $\pmb{a}$. The mask generator is characterized by a 1D-convolution that takes the channels from $B_a$ to $n_{spk}\times C_a$. We then apply an output gate, which involves the element-wise multiplication of the $Tanh$ and sigmoid $\sigma$ activations of two convolutions,
\begin{equation}
\begin{split}
    \pmb{X} &= \mathrm{ReLU}\left(\mathrm{Conv}\left(\mathrm{PReLU}(\pmb{r})\right)\right),\\
    \pmb{Y} &= \mathrm{Tanh}\left(\mathrm{Conv}(\pmb{X})\right) \odot \sigma\left(\mathrm{Conv}(\pmb{X})\right).
\end{split} 
\end{equation}
Next, we split $\pmb{Y}\in \mathbb{R}^{(n_{spk}\times C_a)\times T_a }$ across the channels into $n_{spk}$ parts to get the masks $\pmb{m}_i\in\mathbb{R}^{C_a\times T_a }$ for each speaker $i$. Finally, we compute
\begin{equation}
    \pmb{z}_i = \pmb{a}\odot \pmb{m}_i
    \quad
    \text{for }
    i \in [1,n_{spk}],
\end{equation}
to get the separated audios of each speaker, as a feature map.

\subsection{Decoder}
\label{sec:decoder}

The decoder mirrors the audio encoder of TDFNet in terms of stride, padding and kernel size, and hence we get the equation
\begin{equation}
    \hat{\pmb{s}}_i = \mathrm{TConv}(\pmb{z}_i) = D_a\left(\pmb{z}_i\right)\quad \text{for }i\in [1,n_{spk}],
\end{equation}
where $\mathrm{TConv}$ is now a 1D transposed convolution that converts the $n_{spk}$ separated feature maps into $n_{spk}$ waveform audio streams.

\section{Audio and Video Sub-Network Structure}
\label{sec:avsubnet_chapter}


We have described the entire pipeline of the proposed TDFNet model without details of the audio and video sub-networks. These modules use a TDANet-like structure\cite{TDANet} and since the design for the audio and video sub-networks is the same, we can drop the $a$ and $v$ subscripts from our notation for simplicity. For more specific details of the TDANet implementation, we refer readers to the original paper \cite{TDANet}. Here we will give a brief overview, focusing on our modifications. 

The input to the audio and video sub-networks: $\pmb{a}_{j-1} + \pmb{a}_{0}$ or $ \pmb{v}_{j-1} + \pmb{v}_{0}$, is first bound in place using a depth-wise convolution, and then converted to a lower ``hidden'' dimension $D$ using another 1D convolution with kernel size 1. This will be the input for the main section of TDANet, which can be categorized into three important phases: 
\begin{enumerate}
    \item the bottom-up down-sampling process
    \item the recurrent operator
    \item the top-down fusion process
\end{enumerate}
as shown in Figure~\ref{fig:TDF_internals}.  





\subsubsection*{Bottom-up Down-sampling}




For kernel size $k$ and stride $s$, we next define the normalized depth-wise convolution,
\begin{equation}
    \mathcal{H}_{k,s}(\pmb{X}) := \mathrm{Norm}\left(\mathrm{DWConv}_{k,s}(\pmb{X})\right).
\end{equation}

The bottom-up down-sampling process uses stacked $\mathcal{H}_{k,2}$ layers to obtain the multi-scale set $\pmb{F}$ with different temporal resolutions: 
\begin{equation*}
    \pmb{F} = \left\{\pmb{F}_0 \in \mathbb{R}^{D\times T}
    ,\dots, 
    \pmb{F}_i \in \mathbb{R}^{D\times \frac{T}{2^{i}}}
    ,\dots, 
    \pmb{F}_q \in \mathbb{R}^{D\times \frac{T}{2^{q}} } \right\}.
\end{equation*}

\subsubsection*{Recurrent Operator}

To extract a global view from these features, we down-sample all elements in the set $\pmb{F}$ to the dimensions of the smallest element using adaptive average pooling. Next, we sum all the down-sampled features to generate the global feature map $\pmb{G}^{in}$:
\begin{equation}
    \pmb{G}^{in}  = \sum_{i=0}^q p(\pmb{F}_i),
    \quad
    \pmb{G}^{in} \in \mathbb{R}^{D\times \frac{T}{2^{q}}}.
\end{equation}
To allow the model to understand the complex relationships between the different time steps, a recurrent operator $\mathcal{R}$ is applied along the temporal dimension of $\pmb{G}^{in}$: 
\begin{align}    
        \pmb{G}^{out} &= \mathcal{R}\left(\pmb{G}^{in}\right), 
        \quad
        \pmb{G}^{out} \in \mathbb{R}^{D\times \frac{T}{2^{q}}}.
\end{align}      
The recurrent operator refers to a sequence modelling structure such as a transformer, recurrent neural network (RNN), long-short-term-memory network (LSTM\cite{lstm}) or gated recurrent unit (GRU\cite{gru}).

\begin{figure}[ht]
    \centering    \includegraphics[width=.95\linewidth]{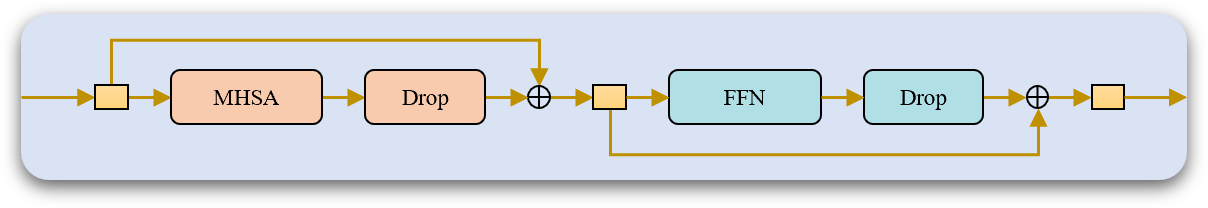}
    \caption{Attention mechanism. This is a diagrammatic view of Equation \ref{eq:inter_processing}.}
    \label{fig:processing_block}
\end{figure}

For the transformer (see Figure \ref{fig:processing_block}), we first use multi-head self-attention (MHSA)\cite{vaswani2017attention_is_all_you_need} followed by a drop block layer\cite{dropblock} and a feed forward network (FFN). The FFN consists of three stacked 1D convolutional layers with kernel sizes \{$1$,  $k$, $1$\}, and number of channels \{$D$, $2D$, $D$\} respectively. Residual connections are added at each stage, so for the transformer the recurrent operator $\mathcal{R}$ is defined as:
\begin{equation}
\label{eq:inter_processing}
    \begin{split}
        \pmb{G}^{mid} &= \mathrm{MHSA}(\pmb{G}^{in}) +\pmb{G}^{in} ,
        \\
        \pmb{G}^{out} &= \mathrm{FFN}\left(\pmb{G}^{mid}\right) +  \pmb{G}^{mid}.
\end{split}
\end{equation}

For the other recurrent operators (RNN, LSTM and GRU) we remove the FFN and the Drop Block\cite{dropblock} layers, but keep the residual connection. We use bidirectional RNNs with the hidden dimension set to the input dimension $D$, hence we also add dropout and a linear projection layer $\mathcal{P}$ from $2D$ channels back to $D$ channels such that the RNN does not change the dimensions of $\pmb{G}^{out}$. This alters Equation \ref{eq:inter_processing} to:
\begin{equation}
        \pmb{G}^{out} = \mathcal{P}\left(\text{RNN}\left(\pmb{G}^{in}\right)\right) +\pmb{G}^{in}.
\end{equation}

\subsubsection*{Top-Down Fusion}

\begin{figure}[ht]
    \centering
    \includegraphics[width=.95\linewidth]{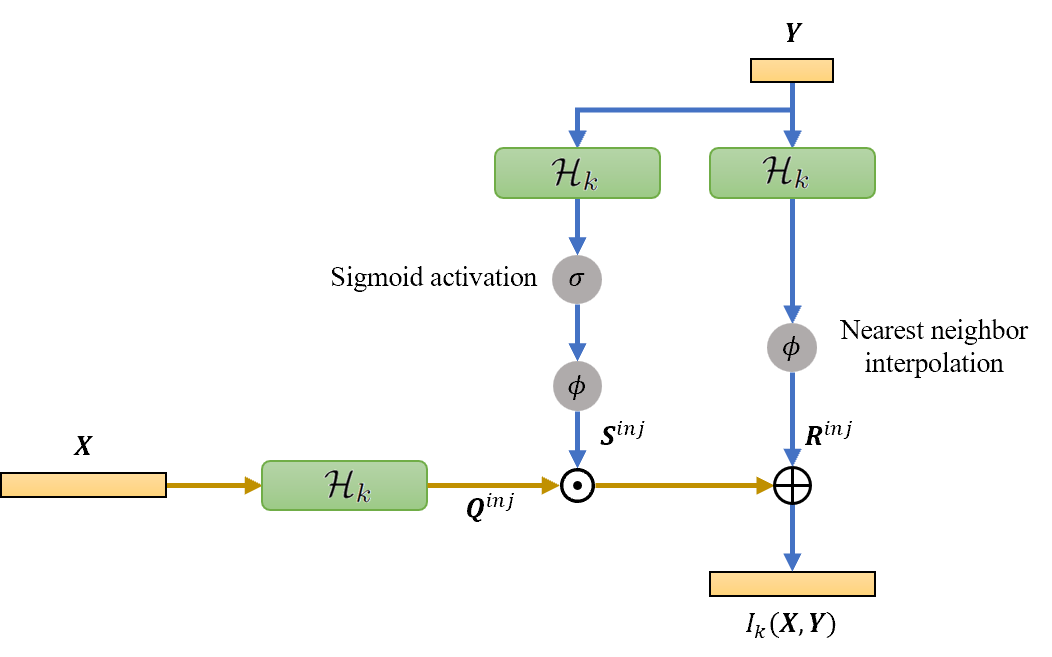}
    \caption{Architecture for the Injection Sum $I_k$ with kernel size $k$ for two inputs $\pmb{X}$ and $\pmb{Y}$ with different temporal dimensions.}
    \label{fig:inj_sum}
\end{figure}

Let $I_k$ define the Injection Sum \cite{TDANet} with kernel size $k$ (see Figure \ref{fig:inj_sum}). Then top-down fusion is defined in two steps. First, the global information $\pmb{G}^{out}$ is fused with each of the multi-scale local features:
\begin{equation}
    \pmb{F}'_i = I_k(\pmb{F}_i, \pmb{G}^{out}),
    \quad
    \pmb{F}'_i \in \mathbb{R}^{D\times\frac{T}{2^i}}
\end{equation}
for $i\in [0,q]$. Next, this multi-scale global-and-local-fusion is collapsed into a single feature map that has a broad view of the entire input. This is achieved with an iterative process using an injection sum with kernel size 1: 
\begin{equation}
    \begin{split}
    \pmb{F}'' &= I_1(\pmb{F}'_{q-1}, \pmb{F}'_{q}) + \pmb{F}_{q-1}, \\
    \pmb{F}'' &= I_1(\pmb{F}'_{q-i-1}, \pmb{F}'') + \pmb{F}_{q-i-1} \quad \text{for } i\in [1,q-1],
    \end{split}
\end{equation}
where after the last iteration, $\pmb{F}'' \in \mathbb{R}^{D\times T}$. This differs from the original TDANet, as in the original implementation $I_1$ is replaced with a different operation. In addition, we have added residual connections from the multi-scale down-sampled features in order to improve training and create a UNet-like \cite{unet} structure. Both of these changes greatly improved the consistency and effectiveness of TDANet without increasing the parameters significantly. Note that in Figure \ref{fig:TDF_internals} we do not show the residual connections as it would make the diagram too complex.

Finally, the feature map $\pmb{F}''$ is converted from the hidden dimension $D$ back to the bottleneck dimension $B_a$ or $B_v$ using a 1D convolution with kernel size 1, and residual connection is added. This result is denoted $\pmb{X}_j$ for the output of the audio sub-network, and $\pmb{Y}_j$ for the output of the video sub-network, see Section \ref{sec:avsubnets_briefMention}.

\section{Experimental Procedures}
\subsection{Dataset}

Following similar methods \cite{VisualVoice, CTCNet, CaffNet_C} in the field, speech separation datasets were constructed from commonly used audio-visual datasets. The speakers in the test set of these datasets did not overlap with those in the training and validation datasets. The raw audio and video frames were obtained using the FFmpeg tool\footnote{https://ffmpeg.org/}. A sampling rate of 16 kHz was chosen, and the model was trained on two seconds of audio and video, equating to audio vectors of length 32,000 and 50 video frames for the 25 FPS video used. 

The LRS2 dataset contains a collection of BBC video clips and three separate folders for training, validation and testing purposes. In order to adapt this dataset to an audio-visual dataset, different speakers were randomly selected two speakers at a time. Their audio signals were mixed using signal-to-noise rations between -5 and 5 dB. The test set is the same as used in previous works\cite{VisualVoice, CTCNet}. In total, the training set contains 11 hours and the validation set contains 3 hours.

\begin{table}[ht]
    \centering    
    \caption{TDFNet hyperparameters.}
    \begin{tabular}{c|l|l}
    \toprule
    Parameter & Value & Description \\
    \midrule 
    $C_a$ & 512 & The audio mixture embedding dimension \\
    $K_a$ & 21 & The kernel size of the encoder and decoder \\
    $S_a$ & 10 & The stride of the encoder and decoder \\
    $B_a$ & 512 & The audio bottleneck out channel dimension \\
    $B_v$ & 64 & The video bottleneck out channel dimension \\ \midrule
    $D_a$ & 512 & The audio sub-network hidden dimension \\
    $q_a$ & 5 & The audio sub-network up-sampling depth \\
    $k_a$ & 5 & The audio sub-network kernel size \\
    $h_a$ & 512 & The GRU hidden dimension \\
    $R_a$ & 16 & The total number of audio sub-network repeats \\ \midrule
    $D_v$ & 64 & The video sub-network hidden dimension \\
    $q_v$ & 4 & The video sub-network up-sampling depth \\
    $k_v$ & 3 & The video sub-network kernel size \\
    $h_v$ & 8 & The number of attention heads \\
    $R_f$ & 3 & The total number of video sub-network repeats \\
    \bottomrule
    \end{tabular}
    \label{tab:TDFNet_Hyperparameters}
\end{table}

\begin{table*}[ht]
    \centering
    \small
    \caption{Audio sub-network recurrent operator ($R_a=4$, $R_f=1$).}
    \begin{tabular}{ll|llll|ll}
    \toprule
    \multicolumn{1}{c}{\multirow{2}{*}{Model}} & \multicolumn{1}{c|}{$A_v$} & \multicolumn{4}{c|}{LRS2-2Mix} & \multicolumn{1}{c}{Params} & \multicolumn{1}{c}{MACs} \\
    \cmidrule(r){3-6}
    \multicolumn{1}{c}{} & \multicolumn{1}{c|}{Module} & \multicolumn{1}{c}{SI-SNRi} & \multicolumn{1}{c}{SDRi} & \multicolumn{1}{c}{PESQ} & \multicolumn{1}{c|}{STOI} & \multicolumn{1}{c}{(M)} & \multicolumn{1}{c}{(G)} \\ \midrule
    CTCNet\cite{CTCNet} &  & 11.1 & 11.6 & 2.90 & 0.895 & 6.3 & 43.6 \\ \midrule
    TDFNet & RNN & 11.6 & 11.9 & 2.95 & 0.903 & \textbf{3.6} & 13.2 \\
    TDFNet & MHSA & 12.9 & 13.1 & 3.08 & 0.921 & 4.2 & \textbf{12.8} \\
    TDFNet & LSTM & 13.4 & 13.5 & 3.08 & 0.928 & 6.8 & 15.7 \\
    TDFNet & GRU & \textbf{13.6} & \textbf{13.7} & \textbf{3.10} & \textbf{0.931} & 5.8 & 15.0 \\
    \bottomrule
    \end{tabular}
    \label{tab:TDFNet_small_results}
\end{table*}

\begin{table}[ht]
    \centering
    \caption{Video sub-network recurrent operator ($R_a=16$, $R_f=3$). } 
    \begin{tabular}{l|llll|ll}
    \toprule
    \multicolumn{1}{c|}{$\beta_j$} & \multicolumn{4}{c|}{LRS2-2Mix} & \multicolumn{1}{c}{Params} & \multicolumn{1}{c}{MACs} \\
    \cmidrule{2-5}
    \multicolumn{1}{c|}{Module} & \multicolumn{1}{c}{SI-SNRi} & \multicolumn{1}{c}{SDRi} & \multicolumn{1}{c}{PESQ} & \multicolumn{1}{c|}{STOI} & \multicolumn{1}{c}{(M)} & \multicolumn{1}{c}{(G)} \\ \midrule
    GRU & 15.3 & 15.4 & 3.19 & 0.942 & 6.6 & 47.2 \\
    MHSA & \textbf{15.8} & \textbf{15.9} & \textbf{3.21} & \textbf{0.949} & \textbf{6.5} & 47.2 \\
    \bottomrule
    \end{tabular}    \label{tab:video_subnet_recurrent_operator}
\end{table}

\begin{table}[ht]
    \centering
    \caption{Sharing parameters in the audio sub-network ($R_a=4$, $R_f=1$)}
    \begin{tabular}{c|llll|ll}
    \toprule
    Audio & \multicolumn{4}{c|}{LRS2-2Mix} & \multicolumn{1}{c}{Params} & \multicolumn{1}{c}{MACs} \\
    \cmidrule{2-5}
    Shared & \multicolumn{1}{c}{SI-SNRi} & \multicolumn{1}{c}{SDRi} & \multicolumn{1}{c}{PESQ} & \multicolumn{1}{c|}{STOI} & \multicolumn{1}{c}{(M)} & \multicolumn{1}{c}{(G)} \\
    \midrule
    \tikzxmark & 12.3 & 12.4 & 3.01 & 0.914 & 22.9 & 15.7 \\
    \tikzcmark & \textbf{13.4} & \textbf{13.5} & \textbf{3.08} & \textbf{0.928} &\textbf{ 6.8} & 15.7 \\
    \bottomrule
    \end{tabular}
    \label{tab:sharing_params}
\end{table}

\begin{table}[ht]
    \centering
    \caption{Sharing parameters in the video and cross-modal fusion sub-networks ($R_a=16$, $R_f=3$).}
    \begin{tabular}{c|llll|ll}
    \toprule
    \multirow{2}{*}{Shared} & \multicolumn{4}{c|}{LRS2-2Mix} & \multicolumn{1}{c}{Params} & \multicolumn{1}{c}{MACs} \\
    \cmidrule{2-5}
     & \multicolumn{1}{c}{SI-SNRi} & \multicolumn{1}{c}{SDRi} & \multicolumn{1}{c}{PESQ} & \multicolumn{1}{c|}{STOI} & \multicolumn{1}{c}{(M)} & \multicolumn{1}{c}{(G)} \\ \midrule
    \tikzcmark & 15.0 & 15.2 & 3.16 & 0.938 & \textbf{4.2} & 38.6 \\
    \tikzxmark & \textbf{15.3} & \textbf{15.4} & \textbf{3.20} & \textbf{0.943} & 4.9 & 38.6 \\
    \bottomrule
    \end{tabular}
    \label{tab:vide_fusion_shared_not_shared}
\end{table}

\begin{table*}[ht]
    \centering
    \small
    \caption{Comparison with SOTA methods ($R_a=16$, $R_f=3$). The ``-" denoted results not reported in the original papers.}
    \begin{tabular}{l|llll|ll}
    \toprule
    \multicolumn{1}{c}{\multirow{2}{*}{Model}} &
      \multicolumn{4}{|c|}{LRS2-2Mix} &
      \multicolumn{1}{c}{Params} &
      \multicolumn{1}{c}{MACs} \\
      \cmidrule{2-5}
    \multicolumn{1}{c}{} &
      \multicolumn{1}{|c}{SI-SNRi} &
      \multicolumn{1}{c}{SDRi} &
      \multicolumn{1}{c}{PESQ} &
      \multicolumn{1}{c|}{STOI} &
      \multicolumn{1}{c}{(M)} &
      \multicolumn{1}{c}{(G)} \\ \midrule
    uPIT\cite{uPIT}               & 3.6           & 4.8           & -             & -              & 92.7          & -             \\
    SuDORM-RF\cite{sudormrf}         & 9.1           & 9.5           & -             & -              & 2.7           & -             \\
    A-FRCNN\cite{AFRCNN}        & 9.4           & 10.1          & -             & -              & 6.3           & -             \\
    Conv-TasNet\cite{Conv-TasNet}    & 10.3          & 10.7          & -             & -              & 5.6           & -             \\ \midrule
    CaffNet-C\cite{CaffNet_C}     & -             & 10.0          & 0.94          & 0.88           & -             & -             \\
    CaffNet-C*\cite{CaffNet_C}  & -             & 12.5          & 1.15          & 0.89           & -             & -             \\
    Thanh-Dat\cite{Thanh_Dat}      & -             & 11.6          & 3.1           & -              & -             & -             \\
    VisualVoice\cite{VisualVoice}  & 11.5          & 11.8          & 3.00          & -              & 77.8          & -             \\
    CTCNet\cite{CTCNet}     & 14.3          & 14.6          & 3.08          & 0.931          & 7.1           & 167.2         \\ \midrule
    TDFNet-small        & 13.6 & 13.7 & 3.10 & 0.931 & 5.8           & \textbf{15.0}         \\
    
    TDFNet (MHSA + Shared) & 15.0 & 15.2 & 3.16 & 0.938 & \textbf{4.2} & 38.6 \\
    TDFNet-large        & \textbf{15.8} & \textbf{15.9} & \textbf{3.21} & \textbf{0.949} & 6.5           & 47.2         \\
    \bottomrule
    \end{tabular}
    \label{tab:TDFNet_main_results}
\end{table*}

\subsection{Hyperparameter Settings}

Following CTCNet\cite{CTCNet} we use 16 total layers with 3 fusion layers. The other hyperparameters are defined in Table \ref{tab:TDFNet_Hyperparameters}. The first block shows the encoder and decoder hyperparameters. The second block shows the audio sub-network hyperparameters. The third block shows the video sub-network hyperparameters.

For training we used a batch size of 4 and AdamW\cite{AdamW} optimization with a weight decay of $1\times10^{-1}$. The initial learning rate used was $1\times10^{-3}$, but the learning rate value was halved when the validation data set loss did not decrease for 5 epochs in a row. We also used gradient clipping in order to limit the maximum $L_2$ norm of the gradient to 5. Training was left running for a maximum of 200 epochs, but early stopping was also applied. Models were all trained on four servers, three containing 8 NVIDIA 3080 GPUs, and one containing 8 NVIDIA 3090 GPUs. The model code is implemented using PyTorch. 

\subsection{Loss Function}
\label{sec:loss_function}

The loss function used for training is the scale-invariant source-to-noise ratio (SI-SNR)\cite{SI_SNR} between the estimated and original signals $\pmb{s}_i$ and $\hat{\pmb{s}}_i$ respectively for each speaker. SI-SNR is defined as
\begin{equation}
    \text{SI-SNR}(\pmb{s}_i,\;\hat{\pmb{s}}_i) = 10 \log_{10} \left(\frac{||\pmb{\omega}_i \cdot \pmb{s}_i||^2}{||\hat{\pmb{s}}_i - \pmb{\omega}_i \cdot \pmb{s}_i||^2}\right),
\end{equation}
where $\pmb{\omega}_i$ is the result
\begin{equation}
    \pmb{\omega}_i = 
    \frac{\hat{\pmb{s}}_i^T \pmb{s}_i}{\pmb{s}_i^T\pmb{s}_i}.
\end{equation}

\subsection{Evaluation Metrics}

Following recent literature, the scale-invariant signal-to-noise ratio improvement (SI-SNRi) and signal-to-noise ratio improvement (SDRi) were used to evaluate the quality of the separated speeches. These metrics were calculated based on the scale-invariant signal-to-noise ratio (SI-SNR)\cite{SI_SNR} and source-to-distortion ratio (SDR)\cite{SDR}:
\begin{equation}
\begin{split}
    \text{SI-SNRi}(\pmb{x}, \;\pmb{s}_i,\; \hat{\pmb{s}}_i ) &= \text{SI-SNR}(\pmb{s}_i, \;\hat{\pmb{s}}_i) - \text{SI-SNR}(\pmb{s}_i, \;\pmb{x}) , \\
    \text{SDRi}(\pmb{x}, \;\pmb{s}_i,\; \hat{\pmb{s}}_i ) &= \text{SDR}(\pmb{s}_i, \;\hat{\pmb{s}}_i) - \text{SDR}(\pmb{s}_i, \;\pmb{x}) ,
\end{split}
\end{equation}
where 
\begin{equation}
    \text{SDR}(\pmb{s}_i, \;\hat{\pmb{s}}_i) = 10\log_{10}\left( \frac{||\pmb{s}_i||^2}{||\pmb{s}_i - \hat{\pmb{s}}_i ||^2} \right).
\end{equation}
We also consider the number of parameters and the MACs. These metrics are important as they determine the computational complexity and memory requirements of the models. In both the number of parameters and MACs, a lower value is preferable. For completeness, we also provide PESQ\cite{PESQ} and STOI\cite{STOI} for the main results tables. For these evaluation metrics, a higher value indicates better performance.

\section{Results}

\subsection{Ablation studies}

In order to evaluate results faster, experimentation was done using a reduced model setting with 1 fusion layer and 3 audio only layers, totaling 4 layers ($R_f=1$, $R_a=4$). 

\subsubsection{Different Recurrent Operators}

In Table \ref{tab:TDFNet_small_results} we examine the effects of using different recurrent operators in the audio sub-network. The transformer, denoted by MHSA, has a good balance between the number of parameters, computational complexity and model performance. It has the lowest number of MAC operations, making the transformer the most efficient choice for audio separation. 

A traditional RNN outperforms CTCNet\cite{CTCNet} by a significant margin, but lacks the huge performance gains of the transformer. However, we can also see that the RNN model uses significantly less parameters - only 58\% of the parameters used by CTCNet. For these experiments, the RNN, GRU and LSTM models all use $h_a=D_a$, the hidden dimension is equal to the input dimension. We found that increasing the hidden dimension to 2$\times$ the input dimension barely affected performance, and resulted in a huge increase in the number of parameters. 

Moving on to the LSTM\cite{lstm} and GRU\cite{gru} structures, we can see that both offer significant gains over the transformer model, and completely outclass the smaller CTCNet model. As pointed out by Max W. Y. Lam et al.\cite{GALR}, it seems that an RNN based approach is better for audio, which features a high temporal correlation, acoustic signal structure, continuities and sequential nature. Interestingly, even though the GRU model has fewer parameters, it outperforms the LSTM architecture by a significant margin, while also utilizing fewer MAC operations. It appears that for this task, the GRU architecture is the optimal recurrent operator in terms of performance, model size and efficiency. It is worth noting however that the GRU is not without its downsides. The GRU structure does use slightly more memory than the LSTM structure of similar size during training. For readers looking to use larger configurations and who do not care about model size, the LSTM may be the better choice in order to train with a larger batch size. 

Table \ref{tab:video_subnet_recurrent_operator} shows the effect of changing the recurrent operator in the video sub-network, as opposed to the audio sub-network experiments in Table \ref{tab:TDFNet_small_results}. Unlike with the audio sub-network, the MHSA clearly outperforms the GRU operator for this medium. These two tables combined show why it is important we use the GRU for the audio sub-network, and MHSA in the video sub-network.

\subsubsection{Sharing Parameters}

In Table \ref{tab:sharing_params} we experiment with sharing the parameters between the audio layers. Both TDFNet and CTCNet achieve a smaller model size compared to VisualVoice\cite{VisualVoice} by sharing parameters. Specifically, the parameters for the audio sub-network $\alpha_j$ are the same for all $j$, the parameters for the video sub-network $\beta_j$ are the same for all $j$ and the parameters for the cross-modal fusion sub-network $\gamma_j$ are also the same for all $j$. In PyTorch, this structure is realised by instantiating one TDFNet block, and passing the data through this same block $R_a$ or $R_f$ times. In Table \ref{tab:sharing_params} we use $R_f=1$, so there are no additional layers for the video sub-network and fusion model to share parameters with. Hence, we experiment with not sharing the parameters between the audio sub-networks. When we stop sharing the parameters between the layers of the audio sub-networks, we are instantiating a new TDFNet block $R_a$ times which results in a huge increase in model size, as seen in Table \ref{tab:sharing_params}. One might expect this to increase the performance of the model, but we see quite the exact opposite effect. Instantiating new layers results in a large drop in performance. This is likely because parameter sharing allows the TDFNet blocks themselves to act as an RNN-like structure, and so when we stop sharing parameters we lose this important effect. 

Table \ref{tab:vide_fusion_shared_not_shared} shows the effect of sharing parameters, this time in the video and fusion layers. If the ``shared" column has a ``$\tikzcmark$" mark, then the parameters for $\beta_j$ are the same for all $j$, and the parameters for $\gamma_j$ are the same for all $j$: we instantiate one instance of the video and fusion sub-networks, and pass the data through these layers $R_f$ times. We can see that unlike in the audio sub-network, sharing parameters here decreases performance. This is likely due to the nature of the fusion operation. In the first iteration, the video network looks at a pure video feature map generated from the video encoder. In subsequent iterations, there is the combined audio signal and the skip connection. It seems that the model likes to use the subsequent iterations to fuse the different information in different ways, and thus benefits from instantiating separate layers. This comes at the cost of increased parameters, but the increase is small and the performance boost is large. It is also worth noting that the models in Table \ref{tab:vide_fusion_shared_not_shared} use MHSA as the recurrent operator in both the audio and video sub-networks. If model size is of most importance, we can see that we can still achieve over 15dB SDRi and SI-SNRi using only 4.2 million parameters - only 60\% of the parameters used by CTCNet. 

\subsection{Comparison with the state-of-the-arts}

In Table \ref{tab:TDFNet_main_results} we can see how TDFNet compares to the competition. We have chosen the three most interesting results for comparison. 

\begin{itemize}
    \item TDFNet-small is the smaller configuration using only one fusion layer, and three audio only layers. It comes from the last row of Table \ref{tab:TDFNet_small_results}. This model is small with outstanding performance and a very low computational cost.
    \item TDFNet (MHSA + Shared) is the version of TDFNet using MHSA as the recurrent operator in both the audio and video sub-networks, and sharing the video sub-network parameters. It comes from the first row of Table \ref{tab:vide_fusion_shared_not_shared}. This model outperforms the SOTA method CTCNet by a significant margin while using only 60\% the number of parameters, and 25\% the number of MACs.
    \item TDFNet-large is the final full-bodied version of TDANet using the GRU in the audio sub-network, and using three separate instances for the video sub-network. It comes from the second row of Table \ref{tab:video_subnet_recurrent_operator}. This model outperforms all other models across all metrics using only 47.2 billion MACs for two seconds of audio sampled at 16000 Hz. This represents approximately 30\% of the MACs used by CTCNet.  At the time of writing this paper, TDFNet is the new SOTA method in audio-visual speech separation. The SI-SNRi score of 15.8 presents a 10\% increase in performance compared to CTCNet.
\end{itemize} 

\section{Conclusion}

Existing multimodal speech separation models are inefficient and have limitations for real-time tasks. We propose a multi-scale and multi-stage framework for audiovisual speech separation based on TDANet and CTCNet. This model can significantly improve the speech separation performance by fusing features of different modalities several times in the fusion stage and influencing the feature extraction network of the corresponding modalities separately. In addition, we explore the impact of different global feature extraction structures on the performance and find that using GRU for sequence modeling can substantially improve the performance and reduce the model computation. Our experiments show that TDFNet outperforms the current state-of-the-art model CTCNet in several audio separation quality metrics while using only 30\% the number of MACs.

\section*{Acknowledgements}

This work was supported in part by the National Key Research and Development Program of China (grant 2021ZD0200301) and the National Natural Science Foundation of China (grant 62061136001).

\bibliography{refs}

\end{document}